# Early Detection of Cancerous Tissues in Human Breast utilizing Near field Microwave Holography

Vineeta Kumari[1], Aijaz Ahmed[1], Tirupathiraju Kanumuri[2], Chandra Shakher[3], Gyanendra Sheoran[4*]

[1]Department of Electronics and Communication Engineering, NIT Delhi, Delhi, India-110040

[2]Department of Electrical and Electronics Engineering, NIT Delhi, Delhi, India-110040

[3]Instrument Design Development Centre, IIT Delhi, Delhi, India-110016

[4]Department of Applied Sciences, NIT Delhi, Delhi, India-110040

**\*Corresponding Author:** sheoran.iitd@gmail.com, gsheoran@nitdelhi.ac.in

**Abstract:**

This work demonstrates an application of near field indirect microwave holography for the detection of malignant tissues in human breast in an effective way. The holograms are recorded by two directive antennas aligned along each other's boresight while performing a raster scan over a 2D plane utilizing XY-linear motorized translation stage and a uniform reference wave. The whole information i.e. amplitude and phase of an object has been provided by indirect holography at microwave frequencies. The extracted phase values are used to determine the dielectric permittivity values which are further utilized for the identification and validating the positions of malignant tissues in the breast phantom. The experimental evaluations performed on the in-house designed and developed tissue mimicking 3D printed breast phantoms. The experimental results demonstrate the ability of microwave holography using directive antennas in locating and identifying the tumors upto the minimum size of 4mm and maximum depth of 25mm in fabricated phantom. The preliminary results present the potential of the Near Field Indirect Holographic Imaging (NFIHI) in order to develop an efficient and economical tool for breast cancer detection.

## 1. Introduction:

Breast cancer is a life-threatening medical emergency which is very common in women. Early stage detection and medical intervention can significantly increase the survival rate of the patients. Although X-Ray digital mammography and MR imaging are the most commonly used and effective techniques for breast screening. Digital mammography is highly ionizing and itself can cause cancer. Moreover, X rays possess ionizing properties, which pose a restriction on the repetitive screenings. Such limitations of standard methods motivate researchers to study and introduce new techniques for cancer detection in breast tissues [1].

Microwave imaging of malignant/benign tissues propounds as an alternative methodology to detect cancerous tissues and it is capable to penetrate up to ample depths beneath the skin tissues, depending upon the tissue compositions [2]. It investigates the dielectric contrast in between the healthy and malignant tissues. Several microwave imaging techniques which are particularly focused on early stage cancer detection have been well documented in the literature, which includes radar imaging, tomography and holographic methods [3].

Amongst all, microwave holography is contemplated as the most effective technique because it gives the complete information of lesions i.e. both amplitude and phase analysis. Microwave holography is categorized as - direct and indirect. In direct holography, the complex scattered field is measured over a selected aperture in the near field using Vector Network Analyzer (VNA) to image the objects. A majority of the holographic approaches for cancer detection [4-7] are based on direct holography. This technique is very expensive and computationally complex as due to its mathematical dependency on vector calculations. On the other hand, indirect holography is two-step process: that employs the method of recording and reconstruction of an object. The hologram is acquired in the form of interference pattern produced by the interference of the reference wave [8] and the object wave-front scattered from its surface. Earlier, the Indirect holographic technique has been reported in the detection of malignancies in breast [9]. But, in the aforesaid work a simple breast model with oil, PCB and silicon to replicate the fat, skin and tumor respectively was used with a simple set of open-ended waveguide probe antennas. The tumor size was defined as 4cm in diameter, which motivates the authors to refine the technique utilizing directive and small size antennas and considering a more realistic breast phantom along with the inclusion of a varying shape and sizes of tumors. Initial results using microstrip and directive antennas for the near field indirect holographic system have been reported by the authors in [10, 11] for imaging of metallic objects.

The focus of the work is to extend the holographic setup for the detection of cancerous tissues in breast. The experimental verification of the proposed imaging system is the main prerequisite before its clinical application, therefore the experimental results for detecting the lesions in a phantom are presented. The patient specific 3D printed phantoms have been designed and developed from the MR images of the real women [12]. The phantoms are fabricated with inclusion of raisins as tumors of different sizes to validate the responsiveness of the system. The materials utilized in fabricated phantoms for the experimental study mimic the actual dielectric properties of tissues and also emulated the dielectric contrast in between the malignant and healthy tissues. Since the tumorous cells exhibit a higher dielectric permittivity because of the presence of higher water content therefore these can be effectively located on the basis of their higher intensity in the image. However, the location and identification of tumor on the basis of higher intensity values but it may result in false positives and false negatives. Hence, it is much required to define such a methodology which can validate that the detected intensities belong to the tumor. That's why, the evaluation of the dielectric permittivity is proposed for the precise detection of malignant tissues. The calculation and mapping of the dielectric properties is performed with the help of the reconstructed phase map of the reconstructed holograms.

## 2. Materials and Methods:
### 2.1. Phantom designing:

The anatomically real breast models are drafted and developed using data set of MR images of real patients. The voxels of the MR images with realistic dielectric properties mapped to the voxels. These voxel models depict the real dielectric properties of the patients of varying ages at frequencies ranging from 0.5GHz - 20GHz. The phantoms are simulated in MATLAB environment. These simulated phantoms are fabricated using tissue mimicking materials having properties of similar to the breast tissues [12]. The fabricated phantoms are shown in Fig.1. The portion of the tissue that is 3D printed using Acronytrile Butadene Styrelene (ABS) plastic corresponds to the locations of fat/adipose tissues. While, the voids in the printed model are correspond to the fibroglandular tissues. The voids are filled with mixture of concentration of gelatine (20g) and water (30g) by weight that constructs fibro glandular tissue equivalent which provides a biologically relevant dielectric contrast with the adipose tissue [13].

Since, the given 3D phantom is derived from the MR images of the real patients, it is correlated with the real distribution of breast tissues. The tumors are mimicked by the raisins of four sizes (i.e. 20mm, 10mm, 5mm, 4mm) introduced in the gelatine after soaking overnight in water.

However, the phantoms are designed and developed for a whole aforesaid frequency range but the proposed experiment is conducted in 'X' band. The measurements with frequencies ranging from 6-14 GHz were performed using VNA (Model No. ZNB40, Rohde and Schwartz) utilizing Nicolson-Ross-Weir (NRW) method [14-15]. Authors have already simulated and presented the dielectric properties of adipose and fibroglandular tissues by employing the Debye/Cole-Cole model parameters [12]. The dielectric properties of the fabricated phantom (i.e. ABS and gelatine) and the inclusions (i.e. raisins) were measured before recording the holograms. Fig 2. shows the measured dielectric properties of tissue mimicking materials (ABS, gelatine and water filled raisins) along with the simulated dielectric properties of the breast tissues for the aforesaid frequency band. Table 1 presented the simulated and measured dielectric values of the tissues at the operating frequency i.e. 8.5 GHz. The tissue mimicking material's dielectric properties are in accordance with the simulated values of the tissue properties.

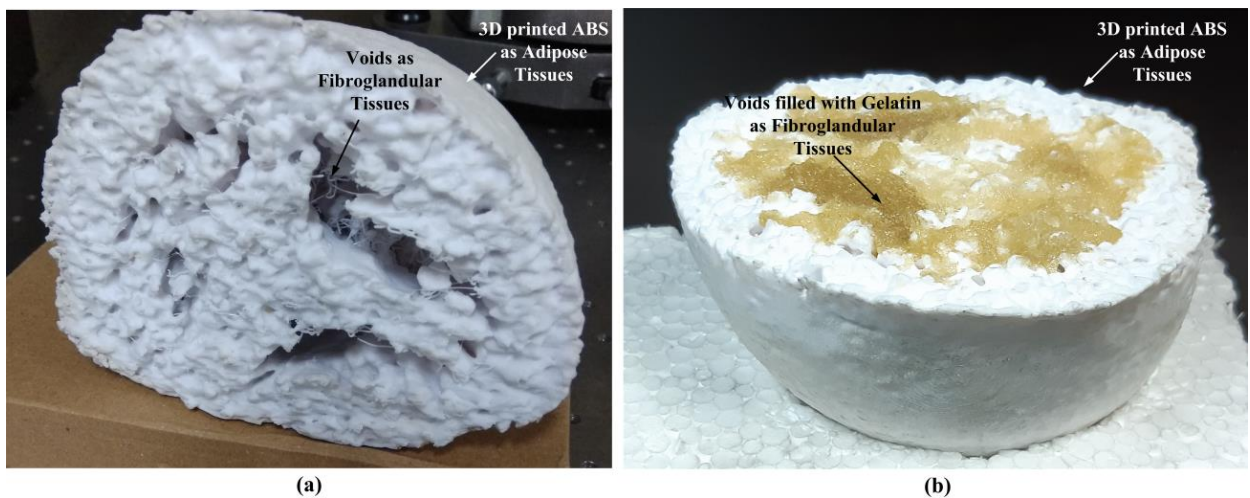

Fig.1: (a) 3D fabricated phantom without filling voids (b) Phantom filled with gelatine.

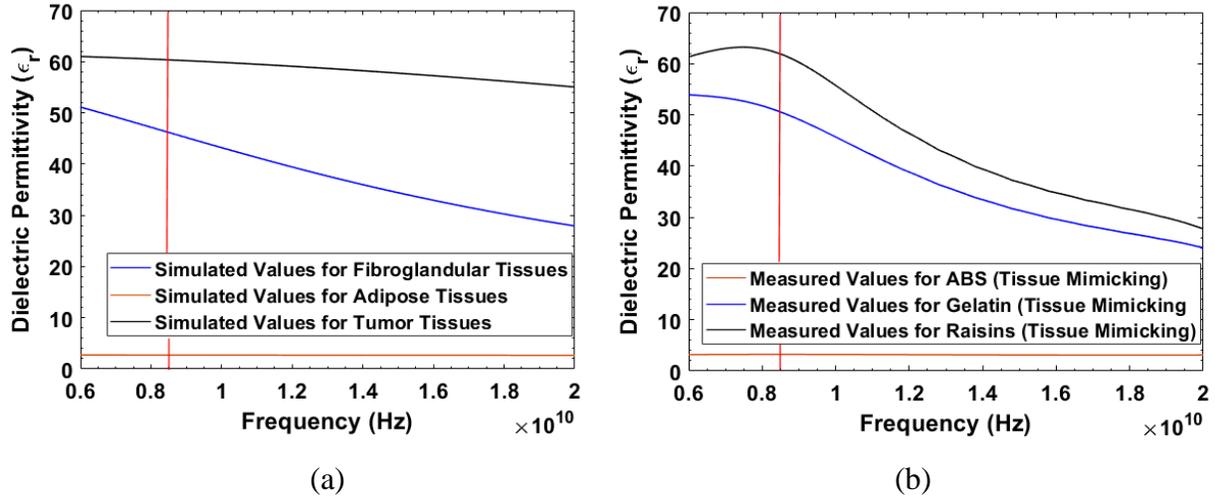

Fig.2: Simulated and measured dielectric values corresponding to (a) breast tissues and (b) tissue mimicking materials respectively.

Table I: Simulated and measured dielectric values at operating frequency (8.5 GHz)

| Tissues | Simulated Values for Tissues | Tissue Mimicking Materials | Measured Values for Tissue Mimicking Materials |
|---|---|---|---|
| Adipose | 2.714 | ABS | 3.248 |
| Fibroglandular | 46.17 | Gelatine | 48.7 |
| Tumor | 60.34 | Raisins | 61.89 |

**2.2. Hologram Recording and Image Reconstruction:** The recording of interference of the object wave and reference wave; and reconstruction of the diffracted field is the basis of the indirect holography. The hologram is recorded while the scattered object wave is combined with the reference wave. The resultant intensity hologram $I_r(x, y)$ in the spatial domain is given by,

$$I_r(x, y) = |S(x, y) \pm R(x, y)|^2 \quad (1)$$

Where, S (x, y) is the scattered field from the object and R (x, y) denotes the reference wave over the x, y plane.

The hologram is reconstructed as a high-quality image by angular spectrum method. In this method, the recorded complex signal's plane wave spectrum in x-y plane is defined by Fourier spectrum of equation (1). The signal is then backpropagated to the object plane and the object is reconstructed by taking the inverse Fourier transform as shown in eqn (2).

$$e(x,y) = FT_{2D}^{-1}\left[ FT_{2D}\left[ \iint I_r(x,y) e^{j(k_x x + k_y y)} dk_x dk_y \right] e^{-jz_0 \sqrt{4k^2 - k_x^2 - k_y^2}} \right] \quad (2)$$

The amplitude and phase of the image are calculated by using equations (3) and (4) respectively,

$$a(x,y) = e(x,y) e^*(x,y) \quad (3)$$

$$\phi(x,y) = \tan^{-1}\left( \frac{im(e(x,y))}{re(e(x,y))} \right) \quad (4)$$

## 3. Experimental Arrangement:

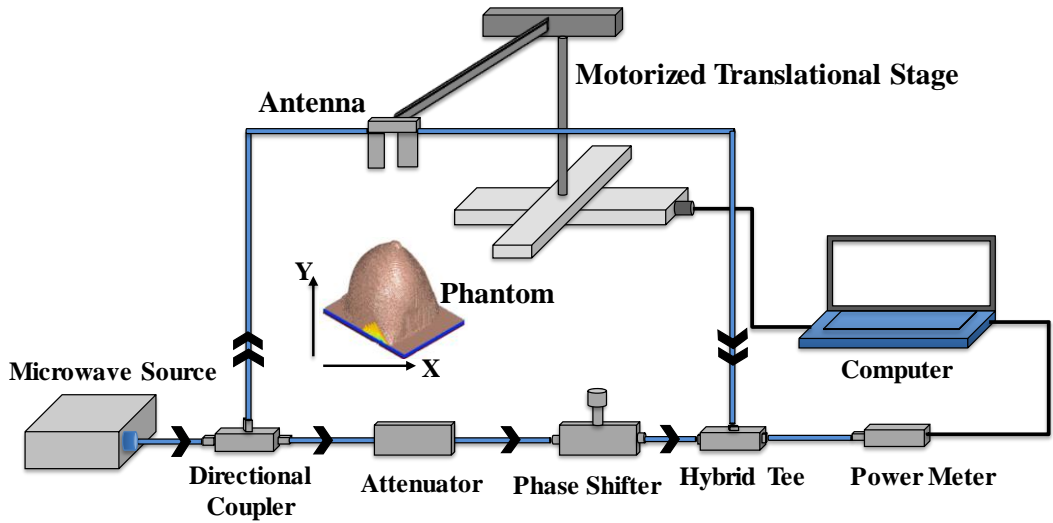

Fig.3: Schematic of the experimental set-up of indirect microwave holography for breast cancer detection.

The schematic of the experimental microwave holographic set-up is demonstrated in Fig. 3. It is used to record the hologram of the fabricated breast phantom with inclusion of tumor. Two Vivaldi antennas with similar designing parameters are used as transmitter and receiver for transmitting and receiving the scattered signals respectively. The details of the antenna design and fabrication are explained in detail in author's previous work [11]. The radiation of both the antennas are framed such that a field directionality of 45º occur to accomplish a better common field of view to avoid mutual coupling. The reference field wave is derived by segregating a portion of the microwave signal utilizing directional coupler. The signal is fed to a variable attenuator, and the signal is attenuated to match the level of the object scattered wave. This is further passed through a phase shifter to provide a linear phase shift to generate the complete phase of 360º in reference wave. The scattered wave from the breast phantom is recorded using the receiving antenna and combined with the reference wave using a magic Tee. The combined output signal intensities are scalarly measured using power sensor, and the resultant intensities are denoted by equation (1). The phantom with outer shell to hold the fabricated adipose tissue and gelatine is oriented in the supine position and the antennas are arranged in the plane using a computer-controlled x-y translation stage to synthesize a 2D planar scanning area as shown in Fig. 4. The experiments are performed without utilizing on any type of matching immersion liquid as a reference hologram without placing the phantoms was recorded to subtract the background reflections.

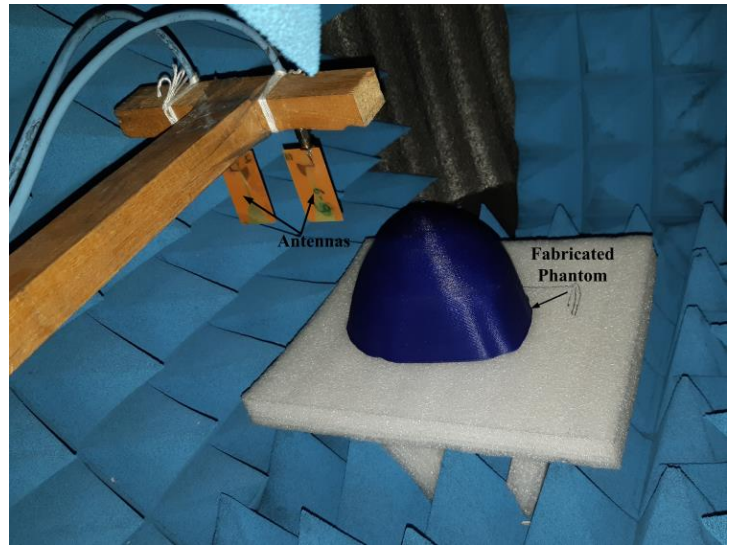

Fig.4: Experimental arrangement of the fabricated phantom and antennas.

The raisins (tumors) are introduced at different positioning levels in the fabricated phantom. Figures 6, 8, 10 described in next section show the top and side view of the breast phantom to depict the positions of the raisins. Instead of considering the shape of tumor as a sphere, realistic shapes of raisins have been taken into consideration.

In order to utilize the antennas for the biomedical applications such as cancer detection, it is a pre-requisite to address the radiation exposure due to the radiated field. A detailed Specific Absorption Rate analysis with all the power levels and locations has been performed to check the effect of antennas radiations on the tissues [16]. Post analysis, it has been validated that the designed antennas possess a safer exposure limit for breast tissues.

## 4. Results and Discussions:

This section presents the experimental results of tumor detection using directive Vivaldi antennas and the 3D phantom. The experiments are performed over a scanning aperture of 20cm × 20cm, at a separation distance of 2.5cm from the object. The spacing of samples is selected as 5mm. The experiments were repeated several times with the inclusion of various sizes of raisins positioned in 3D printed phantom as shown in three experiments in the next section. The repeatability of the experiments is necessary to access the accuracy of the proposed system. Fig. 5 shows the flow chart of the whole process of recording and reconstruction.

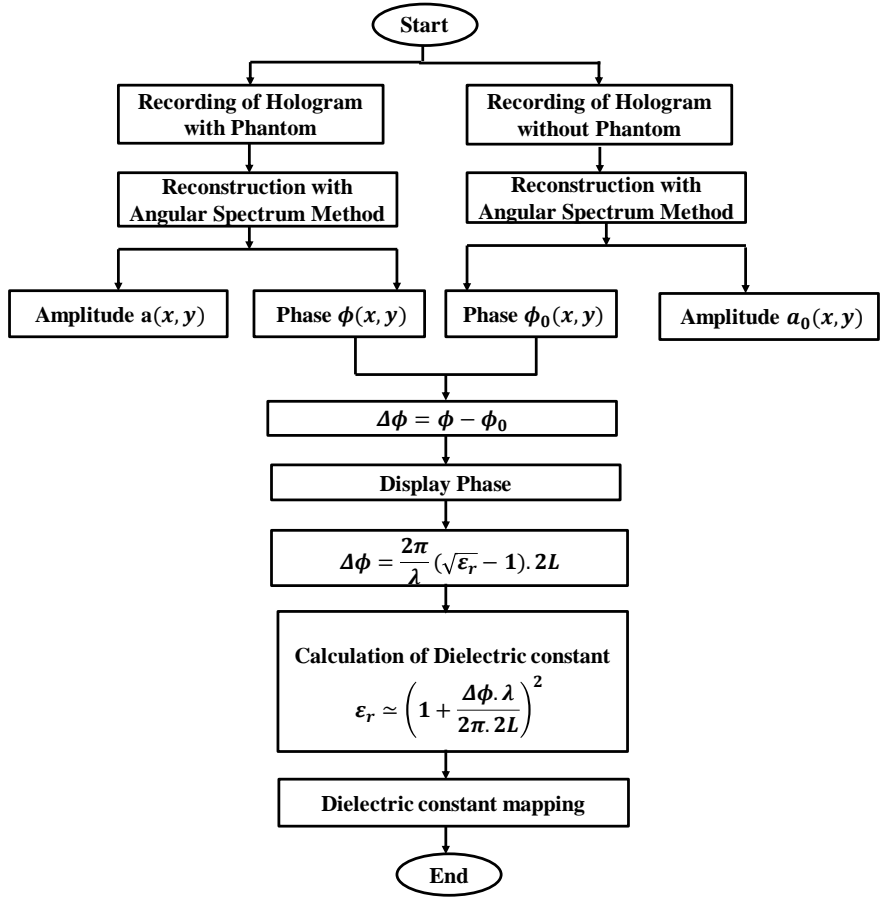

Fig.5: Flow chart for calculating the dielectric permittivity.

To reconstruct an image of the tumor, the recorded hologram is post-processed. The frequency spectrum of the recorded hologram consists of three diffraction orders: DC term, +1 and -1 out of which the required +1 term is filtered out and back propagated on the same plane for reconstruction using angular spectrum diffraction method [10]. The results are presented for tumors of different sizes and shapes in different experiments. The tumors and the phantom outline is marked in white and black colour respectively in all the images.

In first Experiment, two tumors, namely T1 and T2, of sizes 10mm × 10mm and 20mm ×11mm, respectively are introduced in the phantom. The top view and side view of the locations of the tumors are shown in Fig. 6(a) and (b). The amplitude and phase image are displayed in Fig. 7(a) and (b) which clearly displays the inclusions (i.e. raisins) in the reconstructed 2D image at the correct location.

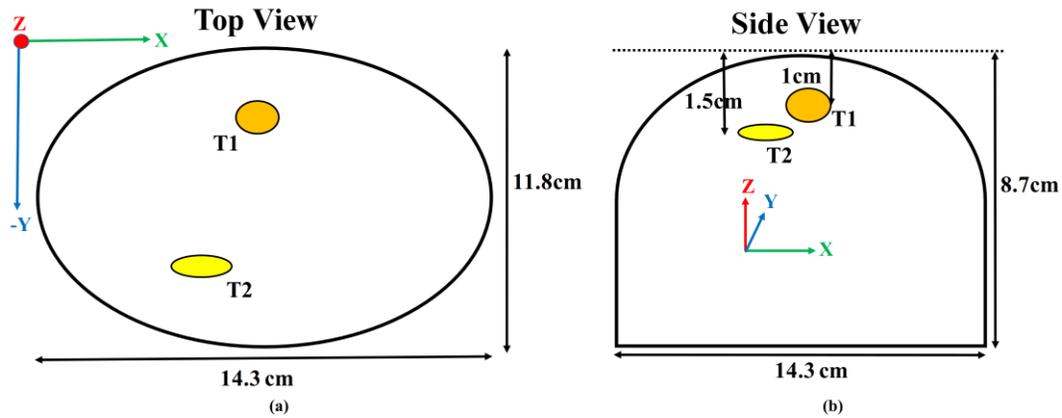

Fig.6: Top and side view of positions of tumors of experiment 1.

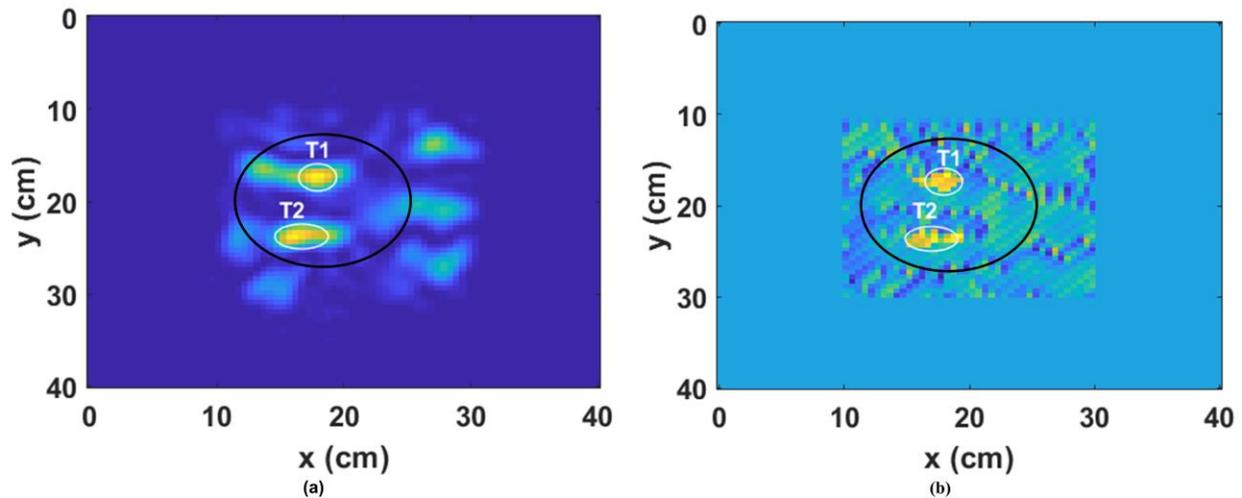

Fig..7: Experiment 1: (a) Reconstructed amplitude image (b) Reconstructed phase image.

In second experiment, Fig. 9 (a) and (b), which corresponds to Fig. 8, clearly shows the highlighted locations of tumors $T_1$ and $T_2$, of sizes 15mm × 10mm and 7mm × 5mm, respectively. The inclusions are clearly shown in the respective amplitude and phase images.

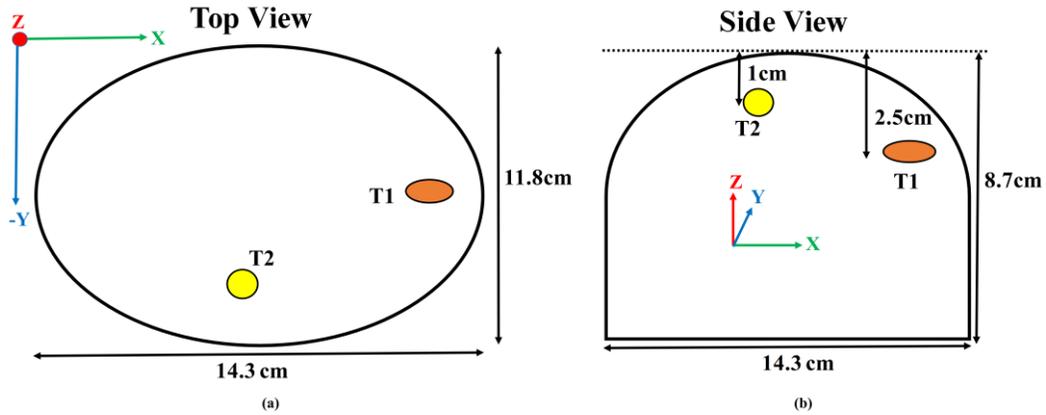

Fig.8: (a) Top and (b) side view of positions of tumors of experiment 2.

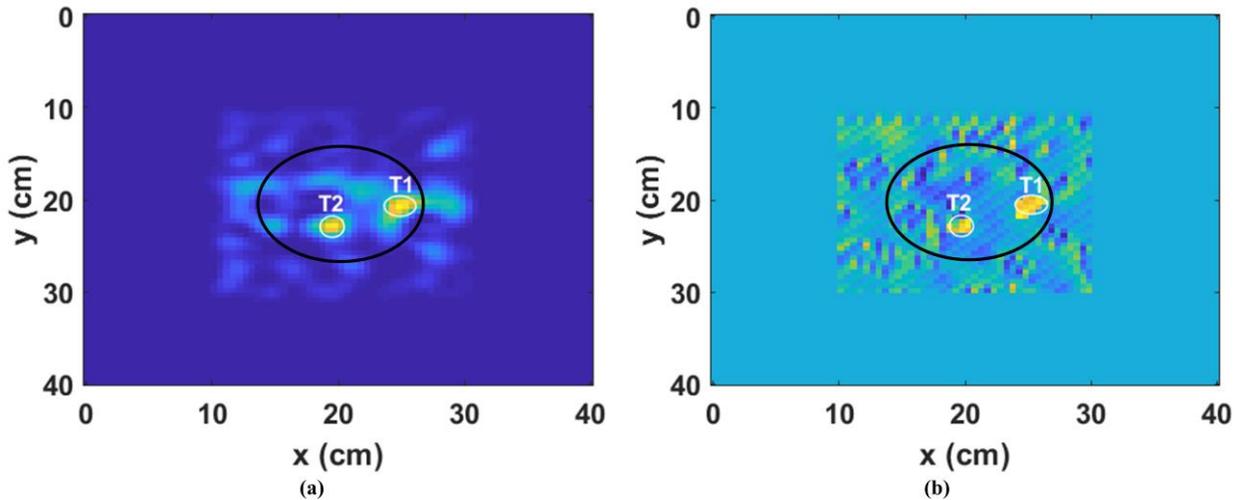

Fig.9: Experiment 2: (a) Reconstructed amplitude image (b) Reconstructed phase image.

In third experiment, Fig. 11(a) and 11(b) show the reconstructed amplitude and phase images corresponding to the experimental arrangement of raisins (tumors) in the phantom as shown in Fig. 10. Five tumors (T1-T5) are included at different heights and positions in the phantom to test the feasibility of the proposed set-up. Figure 10 clearly depicts the location of multiple tumors, $T_1$, $T_2$, $T_3$ and $T_4$ of varying sizes 4mm × 4mm, 7mm × 5mm, 16mm × 14mm, and 6mm × 4mm, respectively. Although, the four tumors are detected and located, but T1 is slightly difficult to be identified in the reconstructed amplitude image due to its small size and less backscattered intensity level as shown in Fig. 11(a). A clearer outline of the T1 can be discerned in the reconstructed phase image i.e. Fig 11(b). Apart from this, T5 cannot be detected because it was located more deeper (i.e. 6.5cm) in the fabricated phantom. This clearly depicts that the tumors around the depth of 2.5 cm are clearly identified and located using the microwave holographic experiments. However, the potential localization of the tumors was done but to validate and identify that the intensity levels actually defines the tumors we have calculated and mapped the dielectric values of the raisins. The mapping of the dielectric properties is performed according to the highly scattered intensity values and corresponding phase values. A detailed mathematical calculation of dielectric values has been presented in the next section along with the resultant images.

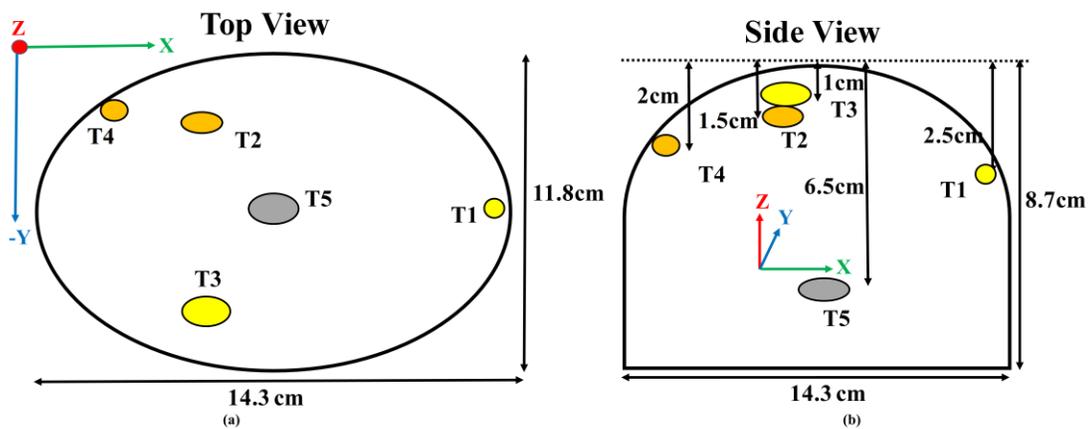

Fig.10: Experiment 3: (a) Top and (b) side view of positions of tumors.

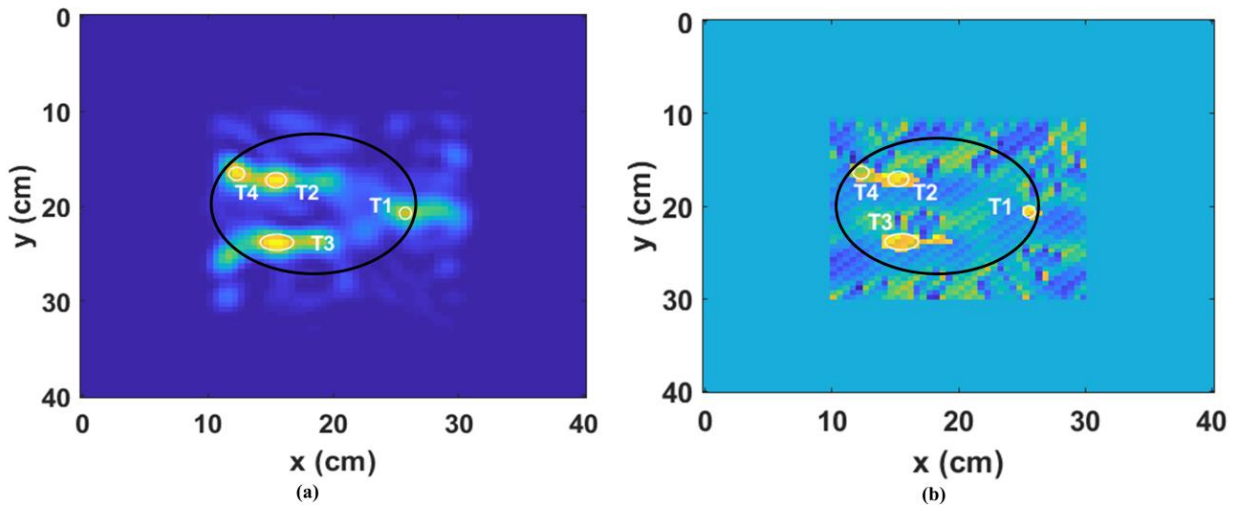

Fig.11: Experiment 3: (a) Reconstructed amplitude image (b) Reconstructed phase image.

**4.1. Dielectric values Calculations:** The identification of tumors has been proposed using the dielectric properties calculation utilizing the phase variations within the objects (i.e. raisins in our case). The phase shift is the difference between the phase angle measured with the object ($\phi$) and phase angle measured without object ($\phi_0$). The reconstructed phase images have the phase variation values that are defined by equation (5)

$$\Delta\phi = \phi - \phi_0 \tag{5}$$

The phase difference is calculated by using the equation (6)

$$\Delta\phi = \frac{2\pi}{\lambda}(n_2 - n_1).2L \tag{6}$$

Where '$n_2$' is the refractive index of the object and '$n_1$' is the refractive index of the air. Now, substitute ($n_1 = 1$) and ($n_2 = \sqrt{\varepsilon_r}$) in equation (6) to obtain,

$$\Delta\phi = \frac{2\pi}{\lambda}(\sqrt{\varepsilon_r} - 1).2L \tag{7}$$

By solving we get,

$$\varepsilon_r \simeq \left(1 + \frac{\Delta\phi.\lambda}{2\pi.2L}\right)^2 \tag{8}$$

Table II: Experimentally calculated Dielectric permittivity values

| S. No. | Tumors | Dielectric Permittivity($\varepsilon_r$) |
|---|---|---|
| Experiment 1 | $T_1$ | 58.65 |
| | $T_2$ | 60.29 |
| Experiment 2 | $T_1$ | 56.46 |
| | $T_2$ | 61.21 |
| Experiment 3 | $T_1$ | 63.24 |
| | $T_2$ | 60.65 |
| | $T_3$ | 59.92 |
| | $T_4$ | 57.17 |

The calculated dielectric values of '$\varepsilon_r$' using equation (8) are shown in Figs. 12 (a-c) and are listed in Table II. Notably, the calculated dielectric values are mapped to the corresponding pixels in accordance with the phase difference. It is clearly observed that these values are in good agreement with the dielectric permittivity given in Table-I. Experimental permittivity values are lying in ±5% of the measured permittivity values of the raisins. The results have effectively validated the potential of the dielectric measurement through phase difference map for identification of the tumors in early stage detection.

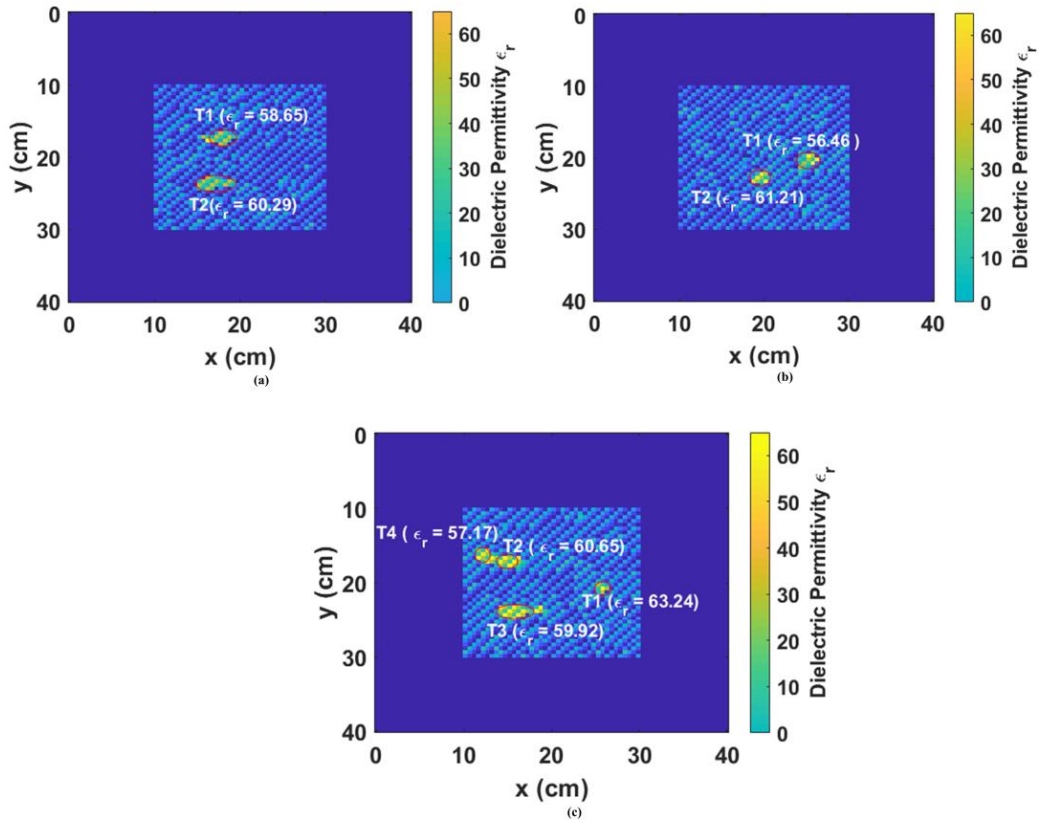

Fig.12: Calculated and mapped permittivity values for experiments 1, 2 and 3 respectively.

**5. Conclusion:** The experimental results of near field indirect holographic technique are demonstrated for early stage detection of lesions in breast tissues utilizing directive antennas. The investigation is performed by making use of the 3D printed phantom with inclusion of raisins to mimic the tumors at various locations and of variable sizes. The reconstructed results of dielectric permittivity values showed that tumors could be detected, located and identified against the background. The detection of lesions size is achieved up to 4mm and a maximum depth of 25mm inside the phantom with an accuracy of ±5%. Here, any matching immersion liquid is not required in between the antennas and breast phantom, which makes the technique very easy and practically implementable. This proposed setup of near field holographic method has the feasibility to become a low cost, compact and easy to use for screening and could be utilize as a diagnostic tool for clinical studies.

**Funding -** This research did not receive any specific grant from funding agencies in the public, commercial, or not-for-profit sectors.

**Disclosure of conflict of interest -** The authors have no relevant conflicts of interest to disclose.

**Research involving Human Participants and/or Animals:**

This manuscript does not contain any studies with human participants or animals performed by any of the authors. Only the MR image database has obtained from the RGCIRC Delhi. The study was submitted to ethics committee of the hospital and was duly approved by them.